\documentclass[aps,twocolumn,showpacs,floats]{revtex4}
\usepackage{graphicx}
\usepackage{dcolumn}
\usepackage{bm}
\usepackage{color}
\usepackage{amssymb}

\begin{document}

\title{Ferromagnetism of a Repulsive Atomic Fermi Gas in an Optical Lattice: A Quantum Monte Carlo Study}

\author{S. Pilati}
\affiliation{The Abdus Salam International Centre for Theoretical Physics, 34151 Trieste, Italy}

\author{I. Zintchenko}
\affiliation{Theoretische Physik, ETH Zurich, 8093 Zurich, Switzerland}

\author{M. Troyer}
\affiliation{Theoretische Physik, ETH Zurich, 8093 Zurich, Switzerland}

\begin{abstract}
Using continuous-space quantum Monte Carlo methods we investigate the zero-temperature ferromagnetic behavior of a two-component repulsive Fermi gas under the influence of periodic potentials that describe the effect of a simple-cubic optical lattice.
Simulations are performed with balanced and with imbalanced components, including the case of a single impurity immersed in a polarized Fermi sea (repulsive polaron).
For an intermediate density below half filling, we locate the transitions between the paramagnetic, and the partially and fully ferromagnetic phases.
As the intensity of the optical lattice increases, the ferromagnetic instability takes place at weaker interactions, indicating a possible route to observe ferromagnetism in experiments performed with ultracold atoms.
We compare our findings with previous predictions based on the standard computational method used in material science, namely density functional theory, and with results based on tight-binding models.
\end{abstract}

\maketitle


Itinerant ferromagnetism, which occurs in transition metals like nickel, cobalt and iron, is an intriguing quantum mechanical phenomenon due to strong correlations between delocalized electrons.
The theoretical tools allowing us to perform \emph{ab-initio} simulations of the complex electronic structure of solid state systems, the most important being density functional theory (DFT)~\cite{parr,hohenberg}, give systematically reliable results only for simple metals and semiconductors. The extension to strongly correlated materials still represents an outstanding open challenge~\cite{cohen}. 
Our understanding of quantum magnetism is mostly based on simplified model Hamiltonians designed to capture the essential phenomenology of real materials.
The first model introduced to explain itinerant ferromagnetism is the Stoner Hamiltonian~\cite{stoner}, which describes a Fermi gas in a continuum with short-range repulsive interactions originally treated at the mean-field level. The Hubbard model, describing electrons hopping between sites of a discrete lattice with on-site repulsion, was also originally introduced to explain itinerant ferromagnetism in transition metals~\cite{hubbard}.
Despite the simplicity of these models, their zero-temperature ferromagnetic behavior is still uncertain.

In recent years, ultracold atoms have emerged as the ideal experimental system to investigate  intriguing quantum phenomena caused by strong correlations.
Experimentalists are able to manipulate interparticle interactions and external periodic potentials independently, allowing the realization of model Hamiltonians relevant for condensed matter physics~\cite{zoller}, or to test exchange-correlation functionals used in DFT simulations of materials~\cite{DFT}.
Indirect evidence consistent with itinerant (Stoner) ferromagnetism was observed in a gas of $^6$Li atoms~\cite{jo09} when the strength of the repulsive interatomic interaction was increased following the upper branch of a Feshbach resonance. However, subsequent theoretical~\cite{pekker} and experimental studies~\cite{ye2012,sanner2012} have demonstrated that three-body recombinations are overwhelming in this regime, and an unambiguous experimental proof of ferromagnetic behavior in atomic gases is still missing.
Proposed modifications of the experimental setup that should favor the reach of the ferromagnetic instability include: the use of narrow Feshbach resonances~\cite{kohstall2012,massignan}, of mass-imbalanced binary mixtures~\cite{conduit2011,ho2013}, reducing the effective dimensionality with strong confinements~\cite{jochim,blume,zinner,conduit2013}, and adding optical~\cite{DFT} and optical-flux lattices~\cite{cooper}.

In this Letter, we use a continuous-space quantum Monte Carlo (QMC)
method to investigate ferromagnetism of a 3D two-component Fermi gas with
short-range repulsive interspecies interactions in the presence of a
simple-cubic optical lattice.
At 3/8 filling (a density of $3/4$ atoms per lattice site) we
obtain the zero-temperature phase diagram as a function of
interaction strength and the amplitude $V_0$ of the optical lattice
focusing on three phases: paramagnet, partially polarized ferromagnet, and fully
polarized ferromagnet.  We do not consider
spin-textured~\cite{conduit09} and antiferromagnetic
phases~\cite{huse,DFT}, nor the Kohn-Luttinger superfluid
instability.

\begin{figure}
\begin{center}
\includegraphics[width=1.0\columnwidth]{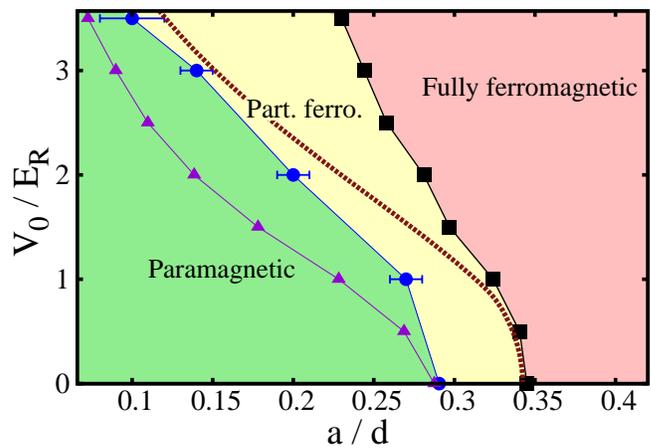}
\caption{Zero-temperature phase diagram at density $nd^3=0.75$, as a function of the interactions strength $a/d$ and the optical lattice intensity $V_0/E_R$. The blue circles separate the region of stability of the paramagnetic phase (green) from the partially polarized ferromagnetic phase (yellow). The black squares separate the partially polarized from the fully polarized ferromagnetic phase (red). The violet triangles and the brown dashed line are the corresponding DFT results. Black and blue lines are guides to the eye.}
\label{fig1}
\end{center}
\end{figure}

Performing simulations in continuous space with an external periodic
potential, rather than employing single-band discrete lattice models
(valid only in deep lattices), allows us to address also the regime of
small $V_0$ and to determine the shift of the ferromagnetic transition
with respect to the homogeneous gas (corresponding to
$V_0=0$)~\cite{conduit09,PRL2010,chang2011,huang2012}. We consider
weak and moderately intense optical lattices, where the noninteracting
band-gap is small or zero.
We find that the critical interaction strength for the transition
between the paramagnetic and the partially ferromagnetic phases (blue
circles in Fig.~\ref{fig1}), as well as the boundary between the
partially and fully polarized ferromagnetic phases (black squares),
rapidly decreases when $V_0$ increases. These results strongly support
the idea of observing itinerant ferromagnetism in experiments with
repulsive gases in shallow optical lattices~\cite{dft_ferro}.
A similar enlargement of the ferromagnetic stability region was
obtained by means of DFT simulations based on the Kohn-Sham
equations~\cite{kohnsham} with an exchange-correlation functional
obtained within the local spin-density approximation
(LSDA)\cite{DFT,newDFT}.
At large lattice depths and interaction strengths, however, we observe
quantitative discrepancies between QMC calculations and DFT due to the strong
correlations which are only approximately taken into account in DFT
methods. This regime, therefore, represents an ideal test bed to develop
more accurate exchange-correlation functionals for strongly correlated
materials.

This scenario appears to be in contrast with the findings obtained for
the single-band Hubbard model, valid for deep lattices and weak interactions, where QMC
simulations indicate that the ground-state is
paramagnetic~\cite{ceperley} (at least up to filling factor $1/4$) and
stable ferromagnetism has been found only in the case of infinite
on-site repulsion~\cite{sorella,kotliar,kivelson}.  
Since at large optical lattice intensity and weak interactions our results agree with Hubbard model simulations (see Supplemental Material~\cite{supplemental}), 
these findings concerning the ferromagnetic transition indicate that the Hubbard model is not an appropriate description for the strongly repulsive Fermi gas in 
moderately deep optical lattices and that terms beyond on-site repulsion and nearest neighbor hopping play an essential role. It also
suggests that the possibility of independently tuning interparticle
interactions and spatial inhomogeneity, offered by our continuous-space
Hamiltonian, is an important ingredient in explaining itinerant
ferromagnetism.
%

%
%
%
We investigate the ground-state properties of the Hamiltonian
\begin{equation}
H = \sum_{\sigma=\uparrow,\downarrow}
       \sum_{i_\sigma=1             }^{N_\sigma      }\left(-\Lambda\nabla^2_{i_\sigma}  + V(\mathbf{r}_{i_\sigma}    )\right)
       +  \sum_{i_\uparrow,i_\downarrow}v(r_{i_\uparrow i_\downarrow}) 
       \;,
\label{hamiltonian}
\end{equation}
where $\Lambda=\hbar^2/2m$, with the atoms' mass $m$ and the reduced Planck constant $\hbar$. The indices $i_\uparrow$ and $i_\downarrow$ label atoms of the two species, which we refer to as spin-up and spin-down
fermions, respectively. The total number of fermions is $N=N_\uparrow+N_\downarrow$, and $r_{i_\uparrow i_\downarrow} = \left|\mathbf{r}_{i_\uparrow}-\mathbf{r}_{i_\downarrow}\right|$.
$V(\mathbf{r})=V_0\sum_{\alpha=x,y,z}\sin^2\left(\alpha\pi/d\right)$ is a simple-cubic optical lattice potential with periodicity $d$ 
and intensity $V_0$, conventionally expressed in units of recoil energy $E_R=\Lambda\left(\pi/d\right)^2$.
$v(r)$ is a short-range model repulsive potential. Its intensity is parametrized by the $s$-wave scattering length $a$, which can be tuned experimentally using Feshbach resonances~\cite{chin}. Off-resonant intraspecies interactions in dilute atomic clouds are negligible since $p$-wave collisions are suppressed at low temperature; hence we do not include them in the Hamiltonian.

%

We perform simulations of the ground state of the Hamiltonian~(\ref{hamiltonian}) using the fixed-node diffusion Monte Carlo (DMC) method. The DMC algorithm allows us to sample the lowest-energy wave function by stochastically evolving the Schr\"odinger equation in imaginary time. 
To circumvent the sign problem the fixed-node constraint is imposed, meaning that the many-body nodal surface is fixed to be the same as that of a trial wave function $\psi_T$. 
This variational method provides the exact ground-state energy if the exact nodal surface is known, and in general the energies are rigorous upper bounds which are very close to the true ground state if the nodes of $\psi_T$ accurately approximate the ground-state nodal surface (see, {\it e.g.}, \cite{Reynolds82,foulkes} and the Supplemental Material~\cite{supplemental} for more details ). 
Our trial wave function is of the Jastrow-Slater form
\begin{equation}
\psi_T({\bf R})= D_\uparrow(N_\uparrow) D_\downarrow(N_\downarrow) \prod_{i_\uparrow,i_\downarrow}f(r_{i_\uparrow i_\downarrow}) \;,
\label{psiT}
\end{equation}
where ${\bf R}=({\bf r}_1,..., {\bf r}_N)$ is the spatial configuration vector and $D_{\uparrow(\downarrow)}$ denotes the Slater determinant of single-particle orbitals of the particles with up (down) spin. 
The orbitals are constructed by solving the single-particle problem in a box of size $L$ with periodic boundary conditions, with and without an optical lattice, obtaining Bloch wave functions and plane waves, respectively. We employ the $N_{\uparrow(\downarrow)}$ lowest-energy (real-valued) orbitals for the up (down) spins.
For homogeneous Fermi gases the accuracy of the Jastrow-Slater form
was verified in Ref.~\cite{chang2011} by including backflow
correlations, and we have performed preliminary simulations with
generalized Pfaffian wave functions~\cite{mitas}, finding no
significative energy reduction. In simulations of the ferromagnetic
transition of the infinite-$U$ Hubbard model fixed-node results were
compared against exact released-node simulations~\cite{carleo} finding
excellent agreement. Furthermore, at large $V_0/E_R$ and small $a/d$ (where our continuous-space Hamiltonian~(\ref{hamiltonian}) can be approximated by the Hubbard model) our results precisely agree with those of Ref.~\cite{ceperley} (see~\cite{supplemental}).
These comparisons give us confidence that the
choice of $\psi_T$ in ~(\ref{psiT}) accurately estimates the
ground-state energy.  The Jastrow correlation term $f(r)$ is obtained
by solving the two-body scattering problem in free space with the
potential $v(r)$ and imposing the boundary condition on its derivative
$f^\prime(r=L/2)=0$. With this choice the cusp condition is satisfied.
Since $f(r)>0$, the many-body nodal surface results only from the antisymmetric character of the Slater determinants.
We simulate systems of different sizes, up to $L=6d$ including $N=162$
fermions, and find that finite-size effects are below statistical
error bars if one subtracts the finite-size correction of
noninteracting fermions $E_0(N_\uparrow,N_\downarrow) -
E_0^{\mathrm{TL}}(P)$, where $E_0^{\mathrm{TL}}(P)$ is the ideal-gas
ground-state energy in the thermodynamic limit (TL) at the polarization
$P=(N_\uparrow-N_\downarrow)/(N_\uparrow+N_\downarrow)$~\cite{ceperleyPRE}.
%

\begin{figure}
\begin{center}
\includegraphics[width=1.0\columnwidth]{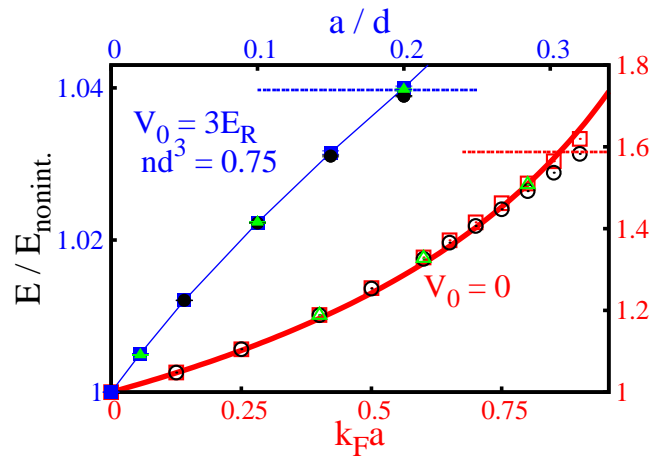}
\caption{Ground-state energy in an optical lattice (upper dataset with full symbols, left and upper blue axes) and in free space (empty symbols, lower and right red axes). Three interatomic potentials are considered: hard spheres (HS, blue and red squares), soft spheres (SS, black circles) and negative power (NP, green triangles). The ranges of interaction strength in the upper and lower $x$-axes coincide if one defines $k_F = (3\pi^2n)^{1/3}$ with the average density $n$ in the optical lattice. The horizontal segments indicate the energies of the fully polarized phases. The thick red curve is the ladder approximation theory for a zero-range pseudopotential~\cite{huang2012}.}
\label{fig2}
\end{center}
\end{figure}

%
%
To model the interspecies interaction, we use prevalently the
hard-sphere potential (HS): $v(r)=+\infty$ if $r<R_0$ and zero
otherwise.  At zero temperature, the properties of a dilute homogeneous
gas are universal and depend only on the two-body scattering
properties at zero energy. These properties are fixed by the $s$-wave
scattering length $a$. For the HS model, one has $a=R_0$.
As $a$ increases, other details of the potential might become
relevant, the most important being the effective range
$r_{\mathrm{eff}}$ and the $p$-wave scattering length
$a_p$~\cite{bishop}, which characterize scattering at low but finite
energy~\cite{delta}.
For homogeneous systems, a detailed analysis of the nonuniversal
effects was performed in
Refs.~\cite{PRL2010,chang2011,huang2012}. Various models with
different values of $r_{\mathrm{eff}}$ and $a_p$ were considered,
including resonant attractive potentials designed to mimic broad
Feshbach resonances with $r_{\mathrm{eff}} \ll n^{-1/3}$ [$n=N/L^3$ is
the density]~\cite{chin}.
In this work we consider the limited interaction regime $k_Fa \lesssim
1$ ($k_F = (3\pi^2 n)^{1/3}$ is the Fermi wave vector), where
differences in the equations of state were found to be marginal (see
Fig.~\ref{fig2}, lower dataset).
In the presence of an optical lattice, the single-particle band
structure further complicates the two-body scattering process.  To
analyze nonuniversal effects in this situation, we compare the
many-body ground-state energies in optical lattices obtained using
three model potentials with the same $s$-wave scattering length: the HS
model;
the soft-sphere potential (SS), $v(r)=v_{\mathrm{SS}}$ if $r<R_0$ and
zero otherwise, with $R_0 = 2a$~\cite{SS}; the negative-power
potential (NP) $v (r)= v_{\mathrm{NP}} /r^{9}$~\cite{NP}.
In Fig.~\ref{fig2} (upper dataset), we show results for an optical
lattice with intensity $V_0=3E_R$. Nonuniversal corrections are found
to be below statistical error-bars up to values of the interaction
parameter where ferromagnetic behavior occurs (see below).
%
%
%
In the following, we use the HS model and parametrize the interaction
strength with the parameters $k_Fa$ and $a/d$, in free space and in
optical lattices, respectively. The latter can be compared with the
former if one defines $k_F$ with the average density in the optical
lattice.

Many theoretical studies of atomic gases in optical lattices have
instead adopted discrete lattice models within a single-band
approximation and with on-site interactions only.
The on-site interaction parameter is usually determined
without considering the strong virtual excitations to higher Bloch
bands which are induced by short-ranged potentials~\cite{buechler}.
This approximation is reliable only if $V_0 \gg E_R$ and $a\ll d$~\cite{jaksch}.
In the regime considered in this work higher-band processes are important
and they can have a strong impact on the properties of discrete-lattice models~\cite{sengstock}.
Reference~\cite{buechler} introduced a different procedure to determine the
on-site Hubbard interaction parameter which is valid at  low filling and effectively takes into account the role of higher
bands.

%
%
\begin{figure}
\begin{center}
\includegraphics[width=1.0\columnwidth]{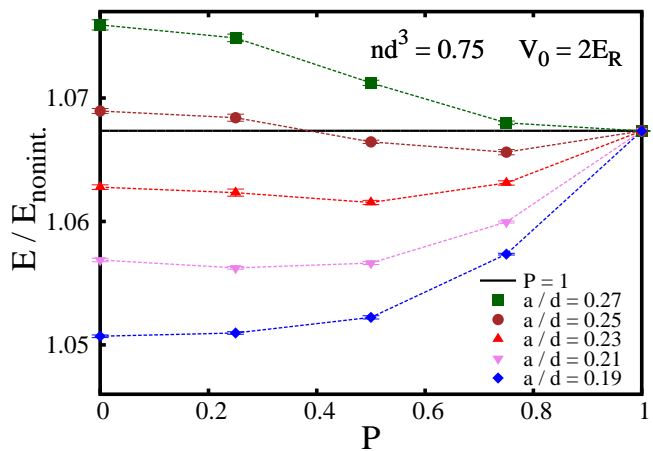}
\caption{Energy versus population imbalance $P=\left(N_\uparrow-N_\downarrow\right)/\left(N_\uparrow+N_\downarrow\right)$ for different the values of interaction strength $a/d$.  The horizontal black line is the energy of the fully polarized gas, dashed lines are a guide to the eye.}
\label{fig4}
\end{center}
\end{figure}
To determine the onset of ferromagnetism using QMC calculations, we perform
simulations of population-imbalanced configurations.  In
Fig.~\ref{fig4}, we plot the energy as a function of polarization $P$
for fixed lattice depth $V_0=2E_R$ and density $nd^3=0.75$ at
different interaction strengths. The minimum of the curve $E(P)$
indicates the equilibrium polarization of ferromagnetic domains. At
the weakest interaction, the minimum is at $P=0$, so the system is
paramagnetic. For larger $a/d$, we observe minima at finite $P$,
allowing us to estimate the critical interaction strength where the
transition to the partially ferromagnetic phase takes place. We do not
investigate here the order of the transition. Our results are
compatible with different scenarios which have been proposed: weakly
first-order~\cite{belitz}, second-order~\cite{huang2012}, or
infinite-order~\cite{carleo} transitions. A similar analysis at
different optical lattice intensities shows that the critical
interaction strength rapidly diminishes as $V_0$ increases (see blue
bullets in Fig.~\ref{fig1}), meaning that the optical lattice strongly
favors ferromagnetism.

%
\begin{figure}
\begin{center}
\includegraphics[width=1.0\columnwidth]{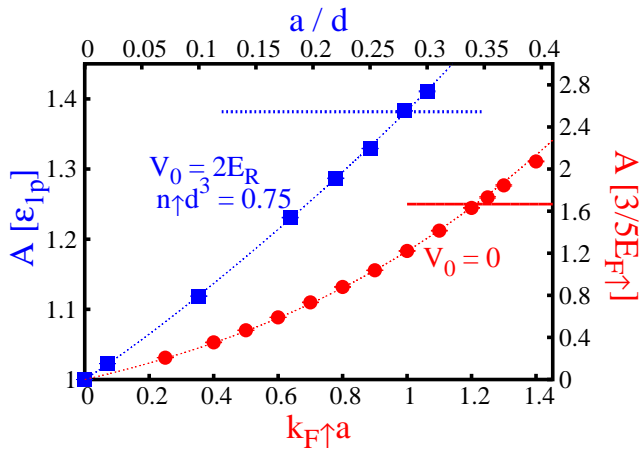}
\caption{(color online).  Chemical potential at zero concentration of the repulsive polaron in an optical lattice (blue squares, left and upper axes) and in free space (red circles, right and bottom axes). $\epsilon_{1p}$ is the energy at the bottom of the noninteracting Bloch band, $E_{F\uparrow}=\hbar^2k_{F\uparrow}^2/2m$. The ranges of interactions strength in the upper and lower x-axes coincide if one defines $k_{F\uparrow}=\left(6\pi^2n\right)^{1/3}$ with the average density $n$ in the optical lattice. The horizontal segments indicate the chemical potential of the majority component.}
\label{fig5}
\end{center}
\end{figure}
 
The critical interaction strength between the partially and the fully
polarized phases is found by considering the problem of the repulsive
Fermi polaron, i.e., a single impurity, say a spin-down
particle, immersed in a fully polarized gas of spin-up particles. In
Fig.~\ref{fig5}, we show the polaron chemical potential $A$,
i.e. the energy of the gas with the impurity minus the energy
of the spin-up particles alone, as a function of the interaction
strength. We compare results obtained in a $V_0=2E_R$ optical lattice
(blue squares), with the homogeneous case $V_0=0$ (red circles, from
Ref.~\cite{PRL2010}). In the region where $A$ is larger than the
chemical potential of the majority component (horizontal segments in
Fig.~\ref{fig5}), the fully polarized phase is stable. By repeating a
similar analysis for different values of $V_0$, the phase boundary
between the two phases (black squares in Fig.~\ref{fig1}) is obtained.
%

In conclusion, we have calculated using QMC methods the ground-state
energy of repulsive Fermi gases in optical lattices as a function of
population imbalance, obtaining the critical interaction strength for
the onset of ferromagnetic behavior.  From simulations of the
repulsive polaron, we determined the region of stability of the fully
polarized phase.
Of particular interest is the question of how effective strongly
correlated single-band models emerge from the continuum description.
In the context of the Mott insulator transition in
bosonic systems, lattice models with only on-site interaction have been compared against
continuous-space simulations, finding for $V_0\gtrsim 4E_R$  only quantitative
differences~\cite{pilati}. However, in the regime of intermediate
values of $V_0$ and strong interactions considered in this work
additional terms such as density-induced tunneling and interaction-induced higher band processes are important, 
and they can induce qualitative changes in the properties of tight binding models~\cite{sengstock,fleischhauer,montorsi}, in particular, concerning the ferromagnetic behavior~\cite{hirsch}.
  These effects are naturally taken into account in a continuous-space description, and our results confirm that they play a role in itinerant ferromagnets.

While in shallow lattices there is good agreement between QMC and
Kohn-Sham LSDA, the regime of deep lattices and strong interactions
represents a new test bed to develop more accurate
exchange-correlation functionals, which is an outstanding open
challenge in material science~\cite{cohen}. Furthermore, our results
show that moderately intense optical lattices are favorable for
experimental realization of ferromagnetism, also due to a faster
thermalization rate compared to very deep lattices.  In a recent
experiment short-range antiferromagnetic correlations have been
observed at half-filling~\cite{esslinger}.

We thank Lianyi He for providing us data from Ref.~\cite{huang2012}, Chia-Chen Chang for the data 
from Ref.~\cite{ceperley},
M. Capone, F. Becca, Lei Wang, and N. Prokof'ev for useful
discussions.  This work was supported by ERC Advanced Grant No. SIMCOFE,
the Swiss National Competence Center in Research QSIT, and the Aspen
Center for Physics under Grant No. NSF 1066293.


\newpage
\begin{center}
\bf{Supplemental Material for\\
 Ferromagnetism of a Repulsive Atomic Fermi Gas in an Optical Lattice: A Quantum Monte Carlo Study}

\author{S. Pilati}
\affiliation{The Abdus Salam International Centre for Theoretical Physics, 34151 Trieste, Italy}

\author{I. Zintchenko}
\affiliation{Theoretische Physik, ETH Zurich, 8093 Zurich, Switzerland}

\author{M. Troyer}
\affiliation{Theoretische Physik, ETH Zurich, 8093 Zurich, Switzerland}
\end{center}


To simulate the ground-state of the many-body Hamiltonian $H$ we employ the Diffusion Monte Carlo (DMC) algorithm.
This technique solves the time-independent Schr\"odinger equation by evolving the 
function $f({\bf R},\tau)=\psi_T({\bf R})\Psi({\bf R},\tau)$ in imaginary time $\tau=it/\hbar$ according to the 
time-dependent modified Schr\"odinger equation
\begin{eqnarray}
-\frac{\partial f({\bf R},\tau)}{\partial\tau}= &-& D\nabla_{\bf R}^2 f({\bf R},\tau) + D \nabla_{\bf R}[{\bf F}({\bf R})
f({\bf R},\tau)] \nonumber \\
&+& [E_L({\bf R})-E_{ref}]f({\bf R},\tau) \;.
\label{FNDMC}
\end{eqnarray}
Here, $\Psi({\bf R},\tau)$ is the many-body wave function while $\psi_T({\bf R})$ denotes the trial function used for 
importance sampling. In the above equation $E_L({\bf R})=
\psi_T({\bf R})^{-1}H\psi_T({\bf R})$ denotes the local energy, ${\bf F}({\bf R})=2\psi_T({\bf R})^{-1}\nabla_{\bf R}
\psi_T({\bf R})$ is the quantum drift force, while $D=\hbar^2/(2m)$ and 
$E_{ref}$ is a reference energy introduced to stabilize the numerics. 
The ground-state energy is calculated from averages of $E_L({\bf R})$ over the asymptotic distribution function $f({\bf R},\tau\to\infty)$. 
While for the ground-state of bosonic systems both $\psi_T({\bf R})$ and $\Psi({\bf R},\tau)$ can be assumed to be positive definite, allowing for the implementation of the diffusion process corresponding to eq.~(\ref{FNDMC}), in the fermionic case the ground-state wave function must have nodes. The diffusion process can still be implemented by imposing the fixed-node constraint $\psi_T({\bf R})\Psi({\bf R},\tau) \geqslant 0$. It can be proven that with this constraint one obtains a rigorous upper-bound of the ground-state energy, which is exact if the nodes of $\psi_T({\bf R})$ coincide with those of the true ground-state~\cite{SReynolds82}. For more details on our implementation of the DMC algorithm, see Ref.~\cite{SBoronatDMC}. 
The DMC algorithm has bee successfully applied to simulate the BEC-BCS crossover in attractive Fermi gases (for a review see Ref.~\cite{Sgiorginireview}), and more recently to investigate the properties of repulsive Fermi gases~\cite{Sconduit09,SPRL2010,Schang2011,Sgora}. It has been extensively applied also to simulate electronic systems with external periodic potentials~\cite{Sfoulkes}.\\
Our trial wave function is of the Jastrow-Slater form
\begin{equation}
\psi_T({\bf R})= D_\uparrow(N_\uparrow) D_\downarrow(N_\downarrow) \prod_{i_\uparrow,i_\downarrow}f(r_{i_\uparrow i_\downarrow}) \;,
\label{psiT}
\end{equation}
where $D_{\uparrow}(N_\uparrow)=\det_{\alpha_\uparrow,i_\uparrow}\left[\phi_{{\alpha_{\uparrow}}}\left({\bf r}_{i_\uparrow}\right)\right]$ denotes a Slater determinant of the up-spin particles, with $\alpha_\uparrow$ an index that labels the $N_\uparrow$ lowest-energy single-particle eigenstates and ${\bf r}_{i_\uparrow}$ the coordinates of particles with up-spin ($i_\uparrow = 1, \dots , N_\uparrow$), while $D_{\downarrow}(N_\downarrow)=\det_{\alpha_{\downarrow},i_{\downarrow}}\left[\phi_{\alpha_{\downarrow}}\left({\bf r}_{i_\downarrow}\right)\right]$ is the Slater determinant of the down-spin particles. $r_{i_\uparrow i_\downarrow} = \left\| {\bf r}_{i_\uparrow} - {\bf r}_{i_\downarrow} \right\|$ denotes the distances between particles with opposite spin.\\
We consider a separable 3D optical lattice of intensity $V_0$ and spacing $d$ with simple-cubic geometry: $V(\mathbf{r}=(x,y,z))=V_0\left[ \sin^2\left(x\pi/d\right) +\sin^2\left(y\pi/d\right)  + \sin^2\left(z\pi/d\right)\right]$.
The single-particle orbitals are constructed by solving the 1D single-particle Schr\"odinger equation:
\begin{equation}
\label{schroedinger}
\left[\frac{-\hbar^2}{2m} \frac{\partial^2}{\partial x^2} + V_0 \sin^2\left(x \pi/d\right)\right]\phi_{q_x}^{(n_x)}(x) = E_q^{(n_x)} \phi_{q_x}^{(n_x)}(x),
\end{equation}
whose solutions are the Bloch functions $\phi_{q_x}^{(n_x)}(x) = \exp\left(iq_x x/\hbar\right)u_{q_x}^{(n_x)}(x)$, with the integer $n_x=1,2,\dots$ being the Band index~\cite{Sashcroft}. In a finite box of size $L=Md$ with periodic boundary conditions ($M$ is a positive integer) the quasi-momentum $q_x \in (-\pi\hbar/d,\pi\hbar/d]$ can take $M$ discrete values. Using the Fourier expansions of the periodic Bloch functions $u_{q_x}^{(n_x)}(x) = \sum_l c_l^{(n_x,q_x)} \exp(i2l\pi x/d)$ and of the optical lattice potential $V(x)=\sum_r V^{(r)} \exp(i2r\pi x/d)$, with the Fourier coefficients $V^{(1)}=V^{(-1)} = -V_0/4$ and $V^{(r)}=0$ if $|r|\neq 1$ (the constant shift $V^{(0)}$ can be set to zero), the Schr\"odinger equation~(\ref{schroedinger}) can be written in matrix form as~\cite{Sbloch}:
\begin{eqnarray}
\sum_l H_{l,l'} \cdot c_l^{(n_x,q_x)} = E_{q_x}^{(n_x)} c_l^{(n_x,q_x)}    ,\;\;\;\mathrm{with}\;  \nonumber \\ 
H_{l,l'} =  \left\{\begin{array}{ll}
(2l+q_xd/(\hbar \pi))^2 E_{\mathrm{R}} &\mathrm{ if }\;\; l=l' \\-V_0/4  & \mathrm{ if }\;\; | l-l' |=1 \\ 0 &\mathrm{ otherwise,} \end{array}\right.
\end{eqnarray}
where $E_\mathrm{R} = \hbar^2\pi^2/(2md^2)$ is the recoil energy. The Bloch functions and the band structure $E_{q_x}^{(n_x)}$ are easily obtained via diagonalization of the matrix $H_{l,l'}$, and we verified that truncating the Fourier expansion beyond $|l| = 5$ one obtains an accurate representation of the Bloch functions in the lowest bands.
The 3D wave functions are given by the products $\phi_{\bf q}^{({\bf n})} = \phi_{q_x}^{(n_x)} \phi_{q_y}^{(n_y)} \phi_{q_z}^{(n_z)}$, with the 3D quasi-momentum ${\bf q} = (q_x,q_y,q_z)$ and the band index ${\bf n} = (n_x,n_y,n_z)$. Pairs of (complex) Bloch functions with opposite quasi-momenta can be replaced by the real-valued combinations $\tilde{\phi}_{\bf q}^{({\bf n})} = (\phi_{\bf q}^{({\bf n})} +\phi_{-\bf q}^{({\bf n})}) / 2$ and $\overline{\phi}_{\bf q}^{({\bf n})} = (\phi_{\bf q}^{({\bf n})} -\phi_{-\bf q}^{({\bf n})}) / 2$. Care must be taken to correctly cut the edges on the Brillouin zone. For $V_0 = 0$, the orbitals $\tilde{\phi}_{\bf q}^{({\bf n})}$ and $\overline{\phi}_{\bf q}^{({\bf n})}$ coincide with the usual free-particle waves: $\tilde{\phi}_{\bf q}({\bf r}) =\cos\left({\bf r}\cdot{\bf q}\right)$ and $\overline{\phi}_{\bf q}({\bf r})=\sin\left({\bf r}\cdot{\bf q}\right)$, with the free-particle momenta ${\bf q} = \frac{2\pi}{L}\left(m_x,m_y,m_z\right)$, where $m_x,m_y,m_z = 0,\pm 1, \pm 2, \dots$. $\tilde{\phi}_{\bf q}$ is used if $m_x > 0$, or if $m_x = 0$ and $m_y > 0$, or if $m_x = m_y = 0$ and $m_z \geqslant 0$, while $\overline{\phi}_{\bf q}$ is used otherwise.\\
As in Refs.~\cite{SPRL2010,Sgiorgini99}, the correlation function $f(r)$ in the Jastrow term (last term in eq.~(\ref{psiT})) is fixed to be the solution of the relative two-particle Schr\"odinger equation in the s-wave channel~\cite{Snewton}:
\begin{equation}
\label{twobody}
\left[-\frac{\hbar^2}{m} \left( \frac{\partial^2}{\partial r^2} + \frac{2}{r} \frac{\partial}{\partial r} \right) + v(r) \right] f(r) = \frac{\hbar^2k^2}{m} f(r).
\end{equation}
The wave-number $k$ is chosen such that $f'(\bar{r}) = 0$, where $\bar{r} \leqslant L/2$ is a matching point used as a variational parameter that we optimize in Variational Monte Carlo simulations, and the normalization is such that $f(\bar{r}) = 1$. We set $f(r) = 1$ for $r > \bar{r}$. While for the hard-sphere and the soft-sphere potentials the analytical solution of eq.~(\ref{twobody}) is known, in the case of the negative-power potential we obtain $f(r)$ numerically using the Runge-Kutta algorithm~\cite{SNR}. The ground-state energy obtained in fixed-node DMC simulations does not depend on the choice of the (positive definite) correlation function $f(r)$, however an accurate choice is useful to reduce the variance.\\

\begin{figure}[htb]
\begin{center}
\includegraphics[width=1.0\columnwidth]{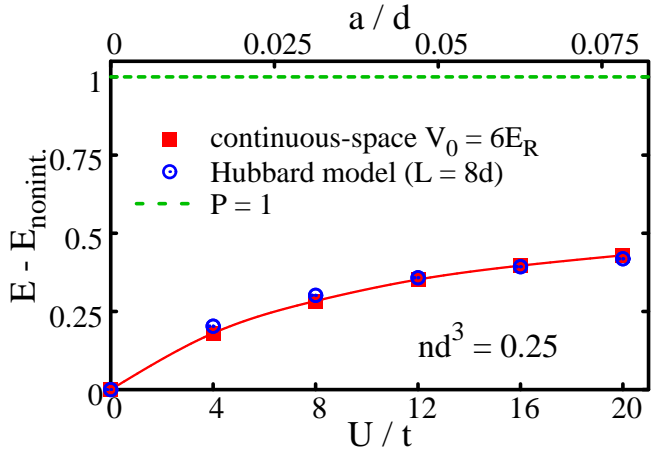}
\caption{Comparison between continuous-space Hamiltonian and Hubbard model. The ground-state energy is plotted as a function of the interaction strength. The offset $E_{\mathrm{nonint.}}$ is the energy of the noninteracting gas at $P=0$. The energy unit is the difference between the fully polarized gas ($P=1$) and $E_{\mathrm{nonint.}}$. The continuous-space interaction parameter $a/d$ is shown in the upper horizontal axis, the Hubbard interaction parameter $U/t$ in the lower axis. Statistical errorbars are smaller tan the symbol-size. The red curve is a guide to the eye.}
\label{figS1}
\end{center}
\end{figure}
%
\begin{figure}[htb]
\begin{center}
\includegraphics[width=1.0\columnwidth]{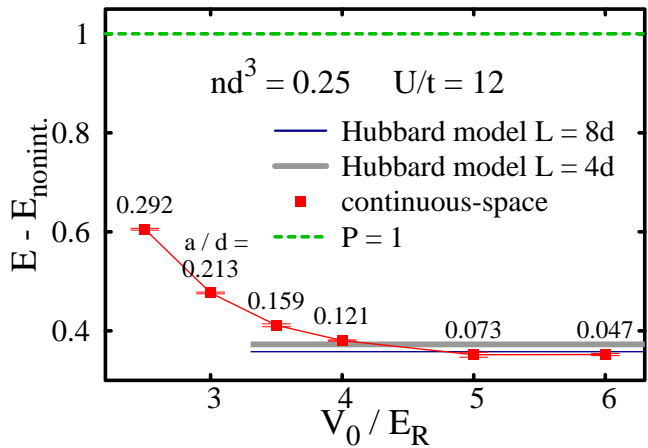}
\caption{Ground-state energy as a function of the optical lattice intensity $V_0/E_R$. The value of scattering length $a/d$ (indicated by the labels placed above to the red squares) is varied in order to fix the Hubbard interaction parameter $U=12t$. The blue and gray bands are the Hubbard model result (the width indicates the statistical errorbar). The dashed green line is the fully polarized phase $P = 1$. Energy offset and units are as in Fig~\ref{figS1}.}
\label{figS2}
\end{center}
\end{figure}
%
%
In order to verify the level of accuracy of the ground-state energies provided by the fixed-node DMC algorithm, we perform a comparison with previous results obtained for the single-band Hubbard model. For deep lattices $V_0/E_R \gg 1$ \emph{and} weak interactions $a / d\ll 1$ ($a$ is the s-wave scattering length), this discrete lattice Hamiltonian is expected to be a reliable approximation of our continuos-space model (eq. (1) in the main text). 
The Hubbard model on a cubic lattice is defined as follows:
\begin{equation}
\label{hubbardmodel}
H = -t \sum_{\left< ij\right> \sigma} \left( c_{i\sigma}^\dagger c_{j\sigma} + \mathrm{H.c.}\right) + U \sum_i n_{i\uparrow}n_{j\downarrow};
\end{equation}
the operator $c_{i\sigma}^\dagger$ ($c_{j\sigma}$) creates (annihilates) one fermion with spin $\sigma$ ($\sigma=\uparrow,\downarrow$), $i$ enumerates the sites in an $N_S = M^3$ lattice, and $\left<ij\right>$ denotes nearest-neighbor pairs. The parameter $t$ is the nearest-neighbor hopping amplitude and $U>0$ is the on-site interaction strength. The ground-state energy of the Hubbard model~(\ref{hubbardmodel}) has been calculated in Ref.~\cite{Sceperley} using the constrained-path Monte Carlo algorithm. For small lattice sizes a benchmark against exact diagonalization results has been performed, finding only minor discrepancies.
The mapping between the parameters of the continuous-space Hamiltonian, namely the optical lattice intensity $V_0/E_R$ and the scattering length $a/d$, to the Hubbard parameters $t$ and $U$ is obtained via a band-structure calculation as outlined in Refs.~\cite{Sjaksch,Sbloch}. Notice that in the conventional procedure to define $U$~\cite{Sjaksch} (which is also adopted here) the interatomic interaction is described by a non-regularized $\delta$~function. This approximation is reliable only for $a/d \ll1$, while for stronger interaction strength virtual excitations to higher bands and the regularization of the potential should be taken into account~\cite{Sbuechler}.
To make a comparison with the Hubbard model we perform continuous-space simulations in a deep optical potential of intensity $V_0 = 6 E_R$, at weak interactions $a/d < 0.1$. The density is set to $nd^3 = (N_\uparrow+N_\downarrow)/M^3= 0.25$ (a value which was considered in Ref.~\cite{S ceperley}). As shown in Figures~\ref{figS1} and~\ref{figS2}, the fixed-node DMC results agree with the results of the constraint-path algorithm. The residual discrepancies are compatible with finite-size effects (DMC data correspond to $M=6$ lattices and include the finite-size correction of the noninteracting system, constraint-path data to $M=4$ and $M=8$ lattices with twisted-averaged boundary conditions). In this regime the unpolarized configurations ($P=(N_{\uparrow}-N_{\downarrow})/(N_{\uparrow}+N_{\downarrow})$ = 0) have much lower energies than the fully polarized states ($P = 1$), indicating a paramagnetic ground state.
Instead, if we diminish $V_0/E_R$ and enlarge $a/d$ (keeping the Hubbard interaction parameter $U/t$ constant) we observe increasing discrepancies between the results of the continuous-space simulations and those performed on the discrete-lattice model (see Fig.~\ref{figS2}). At $nd^3 =0.25$ the deviations from the Hubbard model become evident already in the regime $V_0/E_R \lesssim 4$ and $a/d\gtrsim 0.1$, and are expected to be even more important at the higher density $nd^3 = 0.75$ considered in the main text.\\
The agreement between our results and those of Ref.~\cite{Sceperley} (for large $V_0$ and small $a$) clearly indicates that the ferromagnetic behavior we discuss in the main text is not an artifact of the fixed-node approximation and is instead due to terms, such as density-induced tunneling and higher-band processes, which are not included in the conventional Hubbard model. These terms become important in shallow optical lattices \emph{and/or} at strong interactions.


\begin{thebibliography}{99}
\bibitem{newDFT} We improved the DFT simulations of Ref.~\cite{DFT} using $101^3$ k points.
\bibitem{delta}  The $s$-wave and $p$-wave scattering phase shifts satisfy the relations $k\cot\left(\delta_0(k)\right) = -1/a + r_{\mathrm{eff}}k^2/2 +o(k^4)$ and $k^3\cot\left(\delta_1(k)\right) = -3/a_p^3+o(k^2)$, where $k$ is the scattering wave vector.
For the HS model, one has $r_{\mathrm{eff}}=2a/3$ and $a_p = a$.
\bibitem{NP} For the NP potential: 
$a\cong 0.678\left(v_{\mathrm{NP}} m / \hbar^2 \right)^{1/7}$. 
We determine $r_{\mathrm{eff}}\cong0.63a$ and $a_p\cong1.29a$ by solving numerically the integral equation given in Ref.~\cite{benofy}.
\end{thebibliography}
\end{document}